\documentclass[11pt,nofootinbib,floatfix,onecolumn,preprintnumbers]{revtex4}
\pdfoutput=1
\usepackage{amsmath,amssymb}
\usepackage{graphicx}
\usepackage{rotating}
\usepackage{color}
\usepackage{xcolor}
\usepackage{multirow}
\usepackage{fontenc}
\usepackage{slashed}
\usepackage{longtable}
\usepackage{dcolumn}
\usepackage{bm}
\usepackage{hyperref}
\hypersetup{colorlinks,bookmarksopen,bookmarksnumbered,linkcolor=blue,pdfstartview=FitH,urlcolor=blue,citecolor=blue}

\begin{document}
\title{Deviations to Tri-Bi-Maximal mixing in the limit of $\mu-\tau$ symmetry}

\author{Diana C. Rivera-Agudelo\footnote{Email: diana.rivera11@usc.edu.co}$^{a}$, S. L. Tostado\footnote{Email: sergio.tostado@correounivalle.edu.co}$^{a,b}$, Abdel P\'erez-Lorenzana\footnote{Email: aplorenz@fis.cinvestav.mx}$^{c}$}

\affiliation{\footnotesize
$^{\small a}$ Universidad Santiago de Cali, Facultad de Ciencias B\'asicas, Campus Pampalinda, Calle 5 No. 62-00, C\'odigo Postal 76001, Santiago de Cali, Colombia \\
$^{\small b}$ Departamento de F\'isica, Universidad del Valle, Cll. 13 \# 100 - 0 A.A. 24360, Santiago de Cali, Colombia \\
$^{\small c}$ Departamento de F\'isica, Centro de Investigaci\'on y de Estudios Avanzados del I.P.N., Apartado Postal 14-740, 07000 M\'exico D.F., M\'exico
}

\begin{abstract}
In the limit of an approximate $\mu-\tau$ symmetry in the neutrino mass matrix, we explore deviations to the Tri-Bi-Maximal mixing pattern in the neutrino sector. We consider two different ansatzes for the corrected pattern to predict the current values of neutrino mixing parameters. We show that it is possible to constrain the Majorana $CP$ phases by studying their correlation to the mixing parameters and we study their effects on neutrinoless double beta decay observables. These predictions are sharp for the quasi-degenerate ordering and can be tested in upcoming experiments. 
\end{abstract}

\maketitle

\section{Introduction}\label{sec:intro} 
Neutrino mixing angles have been determined with unprecedent precision in recent years \cite{Tanabashi:2018oca,Capozzi:2013csa,deSalas:2017kay}. These angles define the  structure of the lepton mixing matrix kwown as the Pontecorvo-Maki-Nakagawa-Sakata (PMNS) matrix \cite{Pontecorvo:1957cp,Maki:1962mu}, which can be written in the standard form 
\begin{eqnarray}
\label{PMNS}
U_{\rm PMNS} &=&  \left( \begin{array}{ccc} 
c_{12} c_{13} & s_{12}c_{13} & s_{13} e^{-i\delta_{CP}} \\
- s_{12} c_{23} + c_{12}s_{23}s_{13}e^{i\delta_{CP}}   & 
c_{12}c_{23} + s_{12}s_{23}s_{13}e^{i\delta_{CP}} & -s_{23}c_{13} \\
-s_{12}s_{23} - c_{12}c_{23}s_{13}e^{i\delta_{CP}} & 
c_{12}s_{23} -c_{23}s_{12}s_{13}e^{i\delta_{CP}} & c_{23}c_{13} 
\end{array} \right) \nonumber \\ 
&&\times {\rm diag} \left[ 1,  e^{-i\frac{\beta_1}{2}}, e^{-i\frac{\beta_2}{2}}\right].
\end{eqnarray}
Here, $c_{ij}$ and $s_{ij}$ stand for $\cos \theta_{ij}$ and $\sin \theta_{ij}$, respectively, and $\theta_{ij}$ denotes the mixing angles $\theta_{12}$, $\theta_{13}$, and $\theta_{23}$~. $\delta_{CP}$ is the Dirac-$CP$ violating phase, whereas $\beta_1$ and $\beta_2$ are two additional phases which account for the Majorana nature of neutrinos. Despite the success in determining the mixing angles \cite{Tanabashi:2018oca,Capozzi:2013csa,deSalas:2017kay}, a precise determination of the $CP$ violating phases is still missing. The next-to-next generation of neutrino experiments could finally help to determine the Dirac $CP$ violating phase ($CPVP$) \cite{Abi:2018dnh}. However, given the lack of experimental observables directly related to the Majorana phases, the needed of indirect, but complementary determinations of their values becomes an interesting task to be explored.

Oscillation experiments confirm that the reactor $\theta_{13}$ angle is small but not zero, whereas the atmospheric $\theta_{23}$ angle its close to its maximal value, $\pi/4$. A direct consequence of choosing these critical values ($\theta_{13} =0$ and $\theta_{23}=\pi/4$) is that the mixing matrix in Eq. (\ref{PMNS}), without including Majorana pases, takes the form
\begin{equation}\label{Umutau}
U_{\mu-\tau} =  \left( \begin{array}{ccc} 
c_{12}   & s_{12}   & 0 \\
\frac{-s_{12}}{\sqrt{2}} & \frac{c_{12}}{\sqrt{2}} & \frac{- 1}{\sqrt{2}} \\
\frac{- s_{12}}{\sqrt{2}} & \frac{ c_{12}}{\sqrt{2}}& \frac{1}{\sqrt{2}}
\end{array} \right),
\end{equation}
with the mixing angle $\theta_{12}$ as the only free parameter. The subindex $\mu-\tau$ in Eq. (\ref{Umutau}) refers to the so-called $\mu-\tau$ symmetry \cite{Fukuyama:1997ky,Harrison:2002et,Babu:2002ex,Ohlsson:2002rb,Mohapatra:2004mf,Ghosal:2004qb,Ma:2004zv,Mohapatra:2006un,Joshipura:2005vy,Fuki:2006xw,Riazuddin:2007aa,Luhn:2007sy,Koide:2008zza,Honda:2008rs,Ishimori:2008gp,Ge:2010js,Ge:2011ih,He:2011kn,Ge:2011qn,Talbert:2014bda,He:2012yt} since it satisfies the relations $|U_{\mu i}| = |U_{\tau i}|$. Among the symmetry approaches based on the $\mu-\tau$ symmetry, it has been of great interest the Tri-Bi-Maximal (TBM) mixing pattern \cite{Vissani:1997pa} where $\sin^2 \theta_{12}=1/3$. The TBM pattern has been the starting point of many theoretical works since it can be generated from larger flavor symmetries (see for instance \cite{Altarelli:2012ss,Xing:2015fdg} and references therein). Nonetheless, the predicted mixing angles within this symmetric approach are in conflict with current experimental determinations. 

 Deviations to the TBM scenario have been investigated in order to restore the compatibility with latest neutrino data \cite{Petcov:2014laa,Garg:2013xwa,Garg:2017mjk,Shimizu:2014ria,Sruthilaya2015,Sruthilaya2016}. Some parametrizations written in the form 
\begin{equation}\label{eq:corr}
U_{\rm PMNS} = U_{TBM}U_{\rm Corr}
\end{equation}
have been considered, where $U_{\rm Corr}$ is a correction matrix which encodes the deviation from the TBM mixing matrix $U_{TBM}$. This correction is usually written in terms of up to two rotation angles in the orthogonal case \cite{Garg:2013xwa,Garg:2017mjk}, and including one complex phase in the unitary case \cite{Shimizu:2014ria,Sruthilaya2015,Sruthilaya2016}. These approaches have been able to reproduce the pattern of mixing parameters in the limit of small rotation angles and, in some cases, have provided some predictions for the Dirac $CP$ phase. However, this type of ansatz could not give any hint about the Majorana phases. One possibility of exploring the Majorana case is to include additional phases into the TBM corrected matrix as in \cite{Petcov:2014laa,Chen:2018eou,Chen:2018zbq}. 

On the other hand, the possibility of still having an approximate $\mu-\tau$ symmetry in the neutrino mass matrix has been explored in \cite{Gupta:2013it}, where the viability of such scenario has been confronted with neutrino data. In particular, the connection between an approximate $\mu-\tau$ symmetry in the neutrino mass matrix and both Dirac and Majorana phases have been also explored without considering a particular ansatz of the mixing matrix \cite{Rivera2}. 

In this paper, we investigate different ansatz which correct the TBM mixing and study their effect on the neutrino mass matrix, which, to the best of our knowledge, has not been explored. The novelty of our approach is that, by defining an approximate $\mu-\tau$ symmetry in the mass matrix, the Majorana phases could be bounded and affect neutrino observables as the neutrinoless double beta decay amplitude ($|m_{ee}|$). The remainder of this work is organized as follows. First, we present the different scenarios which correct the TBM pattern. Then, we describe the $\mu-\tau$ symmetric limit of the neutrino mass matrix and its connection with the correction parameters. Third, we show the main results of our analysis and the phenomenological implications over the $CP$ parameters and $|m_{ee}|$. Finally, we give our final comments and conclusions.

\section{Deviations to TBM pattern}

For the sake of simplicity, we will consider in our forthcoming analysis only deviations to the TBM pattern coming from the neutrino sector, such that the corrected mixing matrix takes the form of Eq. (\ref{eq:corr}). Let us now consider the following ansatz for the correction matrix
\begin{equation}\label{Ucorr}
U_{\rm Corr} = U_{ij}(\phi,\sigma) ~\rm{Diag}\left(1,e^{-i\frac{\alpha_1}{2}},e^{-i\frac{\alpha_2}{2}} \right)~,
\end{equation}
where $U_{ij}$ is a unitary matrix which depends on the rotation angle $\phi$, and the complex phase $\sigma$. Here $i,j=1,2,3$, and $i\neq j$. The last term in Eq. (\ref{Ucorr}) is a diagonal matrix with two complex phases, $\alpha_1$ and $\alpha_2$. This approach incorporates in a similar fashion to the standard parametrization in Eq. (\ref{PMNS}) two new complex phases, which are expected to be related to the physical Majorana $CP$ phases. These type of parametrizations, which can be generated in a model dependent way from specific flavor symmetries, has been discussed elsewhere \cite{Petcov:2014laa}.

Among the three possibilities in writing $U_{ij}$, it is direct to see from Eqs. (\ref{eq:corr}) and (\ref{Ucorr}) that only the rotations 
\begin{equation}\label{rot13}
U_{13} = \left( \begin{array}{ccc} 
\cos \phi & 0 & \sin \phi ~e^{-i\sigma} \\
0   & 1 & 0 \\
-\sin \phi ~e^{i\sigma} & 0 & \cos \phi \end{array} \right),
\end{equation}
and
\begin{equation}\label{rot23}
U_{23} = \left( \begin{array}{ccc} 
1 & 0 & 0 \\
0   & \cos \phi & \sin \phi ~e^{-i\sigma} \\
0 & -\sin \phi ~e^{i\sigma} & \cos \phi \end{array} \right) ,
\end{equation}
lead to a mixing matrix able to accommodate $\theta_{13}\neq 0$, as we will show later. We should notice that for recovering the TBM matrix it is enough to consider the limit where $\phi=\alpha_1=\alpha_2=0$. Clearly, in this case, $U_{\rm Corr}$ corresponds to the identity matrix as it is expected. In contrast, however, a $U_{12}$ rotation directly leads to a complex $\mu-\tau$ symmetric matrix predicting $\theta_{13} = 0$ and $\theta_{23} = \pi/4$, with an unphysical Dirac phase. For these reasons, we will not consider the rotation $1-2$ in our following discussion\footnote{In fact, a rotation of the $1-2$ neutrino sector may come from higher order perturbative corrections, which introduces a correction to the zeroth order TBM prediction $\sin^2 \theta_{12}=1/3$. In such case, the predicted mixing matrix still preserves a $\mu-\tau$ symmetric structure.}.


We can see from the standard parametrization of the PMNS matrix in Eq. (\ref{PMNS}) that the mixing angles can be defined in terms of the elements of the neutrino mixing matrix \cite{Tanabashi:2018oca}:
\begin{equation}\label{mixings}
\sin^2{\theta_{12}}=\frac{\mid U_{e2} \mid^2}{1-\mid U_{e3} \mid^2}, ~~~~~\sin^2{\theta_{23}}=\frac{\mid U_{\mu 3} \mid^2}{1-\mid U_{e3} \mid^2}, ~~~~~\sin^2{\theta_{13}}=\mid U_{e3} \mid^2 ,
\end{equation}
while the $CP$ violation parameters can be obtained from
\begin{eqnarray}\label{Jarskog2}
J_{CP} &=& {\rm Im} \left[U_{e1} U_{\mu 2} U_{e2}^{*} U_{\mu 1}^{*} \right]\\
&=&(1-s^2{\theta_{13}})\sqrt{s^2{\theta_{13}}s^2{\theta_{12}}s^2{\theta_{23}}(1-s^2{\theta_{12}})(1-s^2{\theta_{23}})}  \sin\delta_{CP}.  \nonumber 
\end{eqnarray}
By comparing Eq. (\ref{PMNS}) with the corrected $\mu-\tau$ matrix of Eq. (\ref{eq:corr}), we obtain that the Majorana $CP$ phases are related to the new phases via
\begin{equation}\label{phases}
\beta_1 = \alpha_1, \ ~~ \ \beta_2 = \alpha_2 + 2 (\sigma - \delta_{CP}) \ .
\end{equation} 
As we can see, the relations in Eqs. (\ref{mixings}-\ref{phases}) hold independently of the rotation adopted in $U_{ij}$, but may differ when they are written in terms of the correction parameters involved in Eq. (\ref{Ucorr}).  

As a first case, let us consider the $1-3$ rotation (Case I) given in Eq. (\ref{rot13}). From Eq. (\ref{mixings}), the experimental mixing angles are then linked to the correction parameters in Eq. (\ref{eq:corr}) by means of \cite{Shimizu:2014ria,Sruthilaya2015}
\begin{eqnarray}\label{angles}
\sin^2{\theta_{12}} &=& \frac{1}{3-2 \sin^2 \phi}, \nonumber \\
\sin^2{\theta_{23}} &=& \frac{1}{2}\left(1 + \frac{\sqrt{3}\sin 2\phi \cos \sigma}{3-2\sin^2 \phi} \right), \\
\sin^2{\theta_{13}} &=& \frac{2}{3} \sin^2 \phi \nonumber.
\end{eqnarray}
The Jarlskog invariant and the Dirac $CP$ can be obtained from
\begin{equation}\label{CPphase1}
J_{CP} = - \frac{1}{6\sqrt{3}}\sin 2\phi \sin \sigma , \ ~ \ \sin \delta_{CP} = - \frac{(2+\cos 2\phi)\sin \sigma}{\left[(2+\cos 2\phi)^2-3\sin^2 2\phi \cos^2 \sigma\right]^{1/2}}~,
\end{equation}
and the Majorana phases from Eq. (\ref{phases}).

On the other hand, in the second case we take the $2-3$ rotation (Case II) of Eq. (\ref{rot23}). In this case, the mixing parameters are given by \cite{Shimizu:2014ria,Sruthilaya2015}
\begin{eqnarray}\label{angles2}
\sin^2{\theta_{12}} &=& 1-\frac{2}{3 - \sin^2 \phi}, \nonumber \\
\sin^2{\theta_{23}} &=& \frac{1}{2}\left(1 - \frac{\sqrt{6}\sin 2\phi \cos \sigma}{3-\sin^2 \phi} \right), \\
\sin^2{\theta_{13}} &=& \frac{1}{3} \sin^2 \phi \nonumber ,
\end{eqnarray}
while 
\begin{equation}\label{CPphase2}
J_{CP} = - \frac{1}{6\sqrt{6}}\sin 2\phi \sin \sigma , \ ~ \ \sin \delta_{CP} = - \frac{(5+\cos 2\phi)\sin \sigma}{\left[(5+\cos 2\phi)^2 - 24\sin^2 2\phi \cos^2 \sigma\right]^{1/2} } .
\end{equation}
As in the previous case, the Majorana phases can be obtained from Eq. (\ref{phases}).

A direct inspection of Eqs. (\ref{angles}) and (\ref{angles2}) shows that current determinations of the $\theta_{13}$ mixing angle forbid null values of the $\phi$ angle, but are in favor of small values of this parameter. Moreover, this parameter allows determining the size of the departure from the maximal value of the atmospheric {\it via} the reactor angle, where we obtain the approximate sum rules
\begin{eqnarray}
\vert \sin^2 \theta_{23} - \frac{1}{2} \vert &\approx & \sqrt{2} \cos \sigma \sin \theta_{13} ~~~~~~({\rm Case~ I}),\nonumber \\
\vert \sin^2 \theta_{23} - \frac{1}{2} \vert &\approx & \cos \sigma \sin \theta_{13} ~~~~~~ ({\rm Case~ II}).
\end{eqnarray}
Thus, this shows that deviations of $\theta_{23}$ from the maximal mixing are correlated to deviations in $\theta_{13}$ from zero through the phase $\sigma$, which also correlates these mixings with the $CP$ phases, $\delta_{CP}$ and $\beta_2$, as we can see from Eqs. (\ref{phases}), (\ref{CPphase1}) and (\ref{CPphase2}). From the theoretical point of view, this could be of great interest since it point towards the possibility of explaining the observed mixings with a common physical origin in favor of a well-defined flavor symmetry \cite{Rivera1,Rivera2}.

\section{$\mu - \tau$ symmetryc limit in the mass matrix}

From our previous discussion, we can see that it is mandatory to adopt departures from the TBM mixing pattern in order to explain the observed mixings. Nevertheless, it is still missing a description of the effects of such deviations in the neutrino mass matrix. Before doing this, let us first discuss the $\mu-\tau$ symmetry in the mass matrix.

In the basis where the charged lepton are diagonal, the neutrino mass matrix is obtained from  $M_{\nu}=U_{\nu}{\rm diag}(m_1,m_2,m_3) U_{\nu}^{\rm T}$, where $U_{\nu} \equiv U_{\rm PMNS}$. In the $\mu-\tau$ symmetric limit we can write $U_{\nu}=U_{\mu-\tau}$ (as in Eq. (\ref{Umutau})). In this case, the resulting mass matrix reflects an exchange symmetry between the $\mu$ and $\tau$ entries, $|m_{e\mu}|=| m_{e \tau}|~~\text{and}~~ m_{\mu \mu}=m_{\tau \tau}$. It is straightforward to show that the corrections adopted in the mixing matrix will change in consequence this symmetric structure of the neutrino mass matrix.

Following the approach in \cite{Rivera1,Gupta:2013it}, the breaking of the $\mu-\tau$ symmetry in the neutrino mass matrix can be accommodated in two parameters which encode the strength of such breaking. In this way, the mass matrix takes the form
\begin{equation}
M_\nu = M_{\mu - \tau} + \delta M \left(\hat{\delta},\hat \epsilon \right).
\end{equation}
Here, the matrix $M_{\mu-\tau}$ does posses the characteristics of a $\mu-\tau$ symmetric matrix, whereas $\delta M$ is defined by only two nonzero breaking parameters, $\hat{\delta}$ and $\hat \epsilon$. In terms of the matrix elements, these parameters are given by
\begin{eqnarray}\label{deltaepsilon}
\hat \delta &=& 
\frac{ \sum_i (U_{ei} U_{\tau i}-  U_{ei} U_{\mu i} ) m_i}{\sum_i U_{ei} U_{\mu 
i} m_i }~, \nonumber \\
\hat \epsilon &=&
\frac{\sum_i (U_{\tau i} U_{\tau i}-  
U_{\mu i} U_{\mu i}) m_i}{\sum_i U_{\mu i} U_{\mu i} m_i}~.
\end{eqnarray}
It is worth noticing that an approximate $\mu-\tau$ symmetric mass matrix \cite{Gupta:2013it} is obtained when $|\hat \delta| , |\hat \epsilon| \ll 1$. 

It is direct to show that the breaking parameters in Eq. (\ref{deltaepsilon}) can be written in terms of the correction parameters ($\phi,~\sigma$, $\alpha_1$, and $\alpha_2 $) by using Eqs. (\ref{eq:corr}) and (\ref{Ucorr}). The complete expressions of the breaking parameters depend on the parametrization adopted in the mixing matrix and are rather large. Also, they depend on the mass ordering selected and will not be displayed here. The absolute masses $|m_{1,2,3}|$ can be expressed in terms of the lightest neutrino mass $m_0$ as
\begin{eqnarray}
|m_{2}| &=& \sqrt{m_0^2 + \Delta m_{21}^2} ~~,~~ |m_3| = \sqrt{m_0^2 + |\Delta m_{31}^2|}~~ {\rm for~ NH}, \nonumber \\
|m_{1}| &=& \sqrt{m_0^2 + |\Delta m_{31}^2|} ~~,~~ |m_2| = \sqrt{m_0^2 + |\Delta m_{31}^2| + \Delta m_{21}^2}~~ {\rm for~ IH}.
\end{eqnarray}
Here, the square mass difference $\Delta m_{21}^2 = m_2^2-m_1^2$ is also known as the solar mass scale, and $\Delta m_{31}^2=m_3^2 - m_1^2$ as the atmospheric mass scale, where $m_0$ becomes $|m_1|$ for the normal mass hierarchy (NH), and $m_3$ for the inverted mass hierarchy (IH). Thus, the expressions in Eq. (\ref{deltaepsilon}) give us a direct relation between the correction parameters of the mixing matrix and the breaking parameters of the mass matrix. In addition, small departures from the $\mu-\tau$ symmetric limit in the mass matrix restrict the parameter space of $|\hat \delta|$ and $|\hat \epsilon|$ to small values. As a consequence, it could be possible to restrict even more the values of the correction parameters and hence the predicted neutrino mixings.       

\section{Results}
For the numerical analysis, we use the results of the latest global fit for the various neutrino oscillation experiments \cite{deSalas:2017kay}. For the sake of simplicity, we will work hereafter in the inverted hierarchy (IH), however, as we will show later, our results are valid for both approaches. The $3\sigma$ intervals of the mixing angles are
\begin{equation}
0.273<\sin^2 \theta_{12} <0.379,~~  0.0199<\sin^2 \theta_{13}<0.0244,~~ 0.453<\sin^2 \theta_{23}<0.598
\end{equation}
and the squared mass differences $\Delta m_{21}^2 = 7.55^{+0.20}_{-0.16}\times 10^{-5}$ eV$^2$ and $|\Delta m_{31}^2|=2.42^{+0.03}_{-0.04}\times 10^-3$ eV$^2$. Let us divide our discussion into the two cases of interest. First, we analyse the $1-3$ rotation, then, the $2-3$ rotation case.


\subsection{Case I: 1-3 rotation}
Based on Eqs. (\ref{phases}), (\ref{angles}), and (\ref{CPphase1}), we will analyse the impact of the correction parameters ($\phi, \sigma$ and $\alpha_{1,2}$) over the mixing parameters when an approximate $\mu-\tau$ mass matrix is required by imposing $|\hat \delta|,|\hat \epsilon| < 0.25$. We will also analyse the case when this condition is omitted. In any case, we will select those values of the correction parameters which predicts mixing angles within the allowed $3\sigma$ region of the experimental values, and show several relations among the predicted mixings. $\phi$ will be scattered in the ($0,\pi/2$) interval, and the phases $\sigma$, $\alpha_1$, and $\alpha_2$ within the ($-\pi,\pi$) range.

\begin{figure}
\includegraphics[scale=0.6]{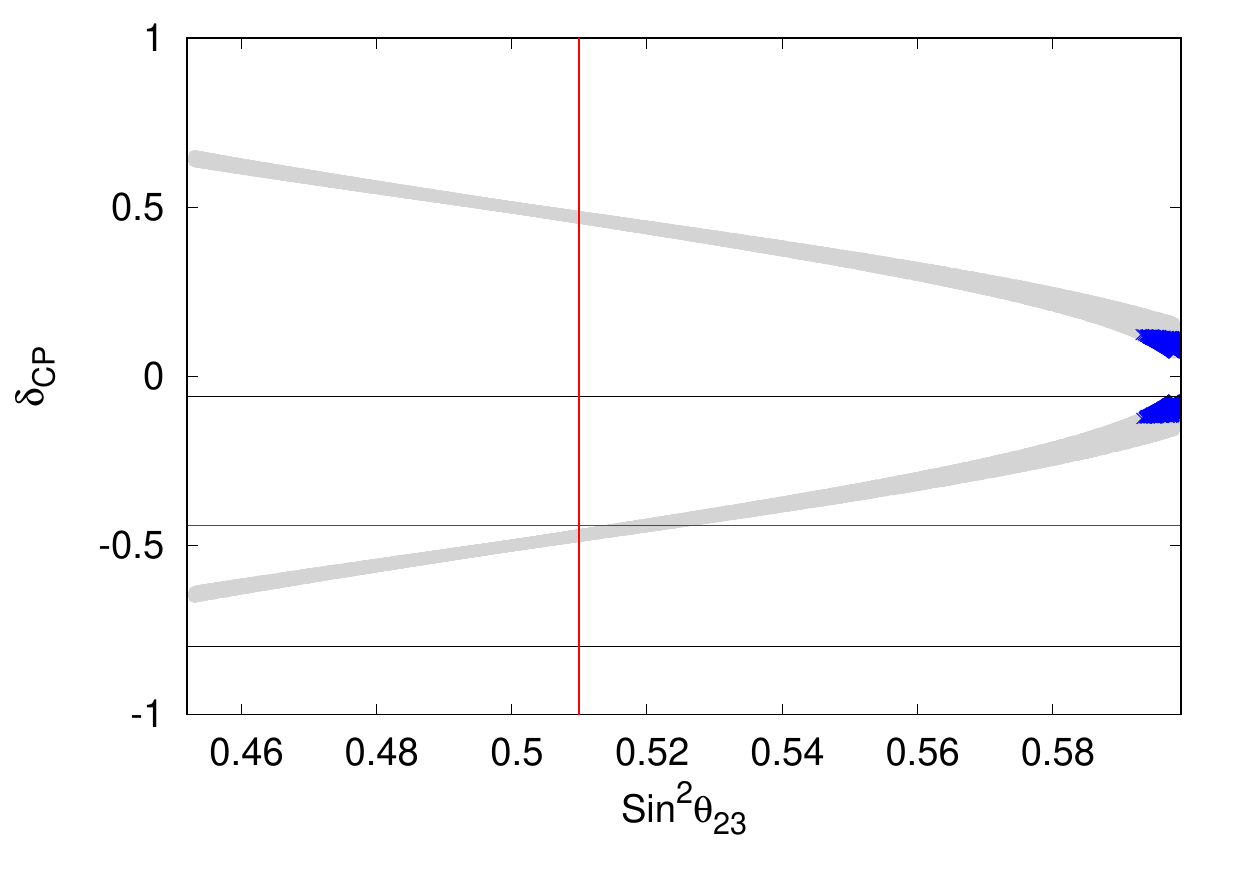} \\
\caption{Correlation between $\delta_{CP}$ and $\sin^2 \theta_{23}$ in Case I. The blue region is obtained for $\alpha_2 =0$, in the limit of an approximate $\mu-\tau$ symmetry. The gray region is obtained when the small breaking requirement is omitted, and does not depend of the chosen values of $\alpha_1$ and $\alpha_2$. Vertical (red) line shows the best fit value of $\sin^2 \theta_{23}$. Horizontal band shows the allowed $3\sigma$ range of $\delta_{CP}$ and its central value (horizontal red line) as indicated by the global fits \cite{deSalas:2017kay}.}\label{fig:CP}
\end{figure}

\begin{figure}
\includegraphics[scale=0.6]{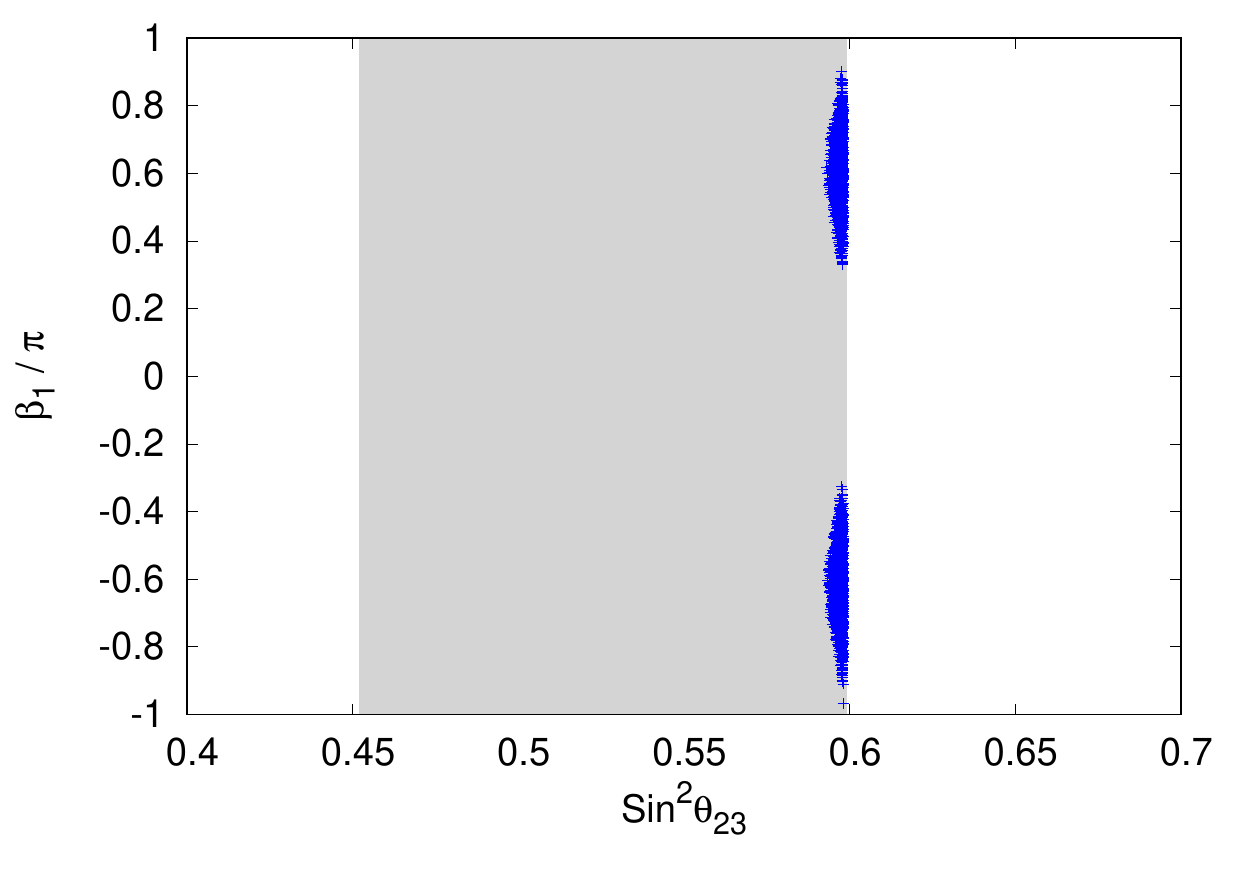}
\includegraphics[scale=0.6]{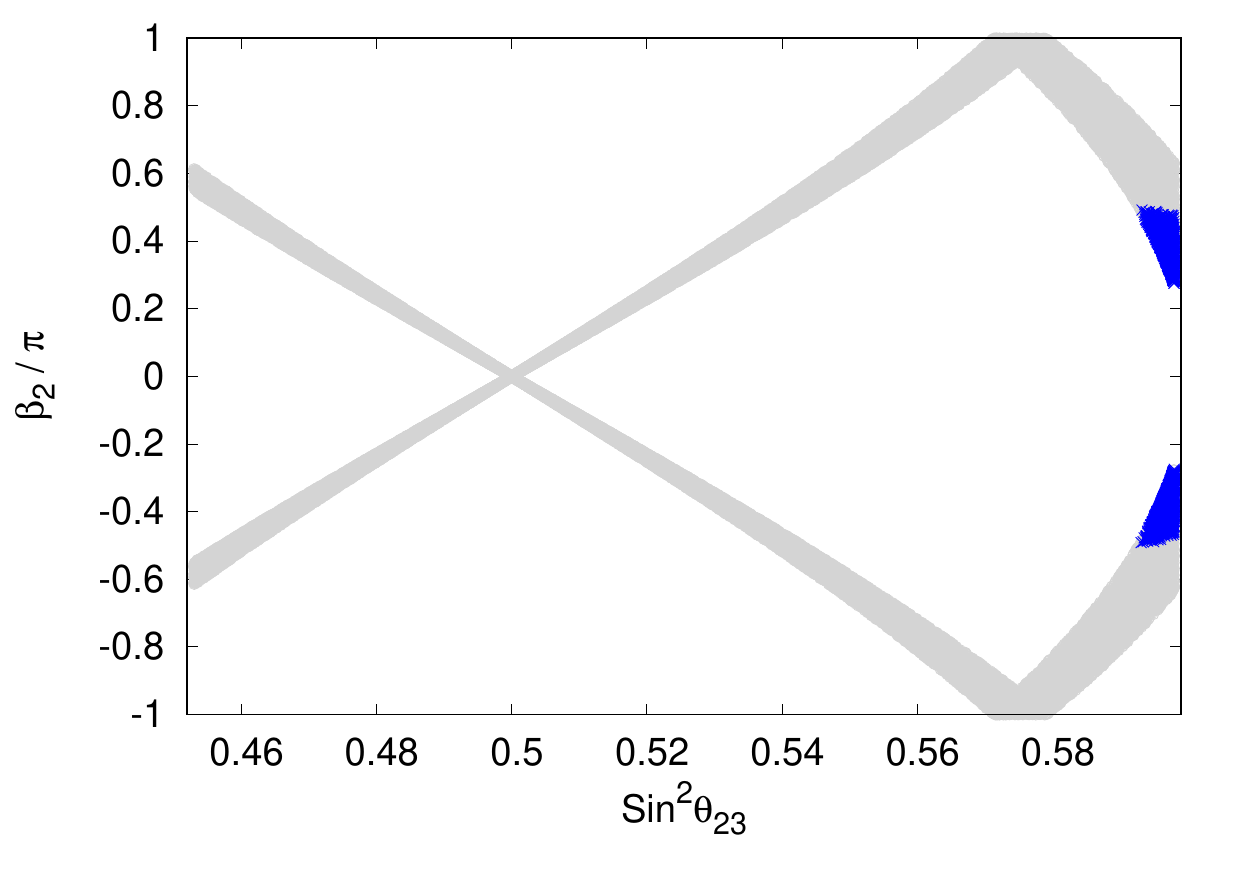}
\caption{Allowed region of Majorana phases in Case I for $\alpha_2 = 0$. Blue (gray) region is obtained when the small $\mu-\tau$ symmetry breaking in the mass mtrix is considered (omitted). Correlations between $\beta_1$ ($\beta_{2}$) and $\sin^2 \theta_{23}$ is shown in the left (right) plot for the $3\sigma$ range.}\label{fig:Majorana}
\end{figure}

Some particular cases can be analysed by selecting specific values of $\alpha_{1}$ and $\alpha_{2}$. Let us first discuss the case $\alpha_2=0$. From Eq. (\ref{phases}), this limit implies that $\beta_1 = \alpha_1$ and $\beta_2 = 2(\sigma - \delta_{CP})$. In Figs. \ref{fig:CP} and \ref{fig:Majorana}, we can observe that the correlations between the $\theta_{23}$ mixing angle and the $CP$ parameters are narrow when both conditions, small symmetry breaking and $\alpha_2=0$, are imposed (blue region). In this case, the deviation of $\theta_{23}$ from its maximal value ($\theta_{23}=\pi/4$) is rather large, i.e. the predicted value lies in a very narrow range near the 3$\sigma$ current limit. A most precise experimental determination of such observable could, in principle, rule out this scenario in the near future. The allowed $3\sigma$ region of $\delta_{CP}$ \cite{deSalas:2017kay} is also shown in Fig. \ref{fig:CP}, which shows that this scenario is also disfavored by the last indications of maximal $CP$ violation ($\delta=-\pi/2$) \cite{Tanabashi:2018oca,Capozzi:2013csa,deSalas:2017kay}. However, as we can see from Figs. \ref{fig:CP} and \ref{fig:Majorana}, there is a strong correlation between $\beta_2$ and $\delta_{CP}$ with the atmospheric angle even when the small breaking condition is omitted (gray region), as expected from Eq. (\ref{phases}). Thus, in this approach it could be possible to extract some indirect information about $CP$ violating phases through a precise determination of the atmospheric angle. Moreover, we can identify that the restricted blue region of $\delta_{CP}$ in Fig. \ref{fig:CP} is due only to the small symmetry breaking conditions since $\delta_{CP}$ is independent of the correction phases $\alpha_{1,2}$, as we can verify from Eq. (\ref{CPphase1}).

\begin{figure}
\includegraphics[scale=0.6]{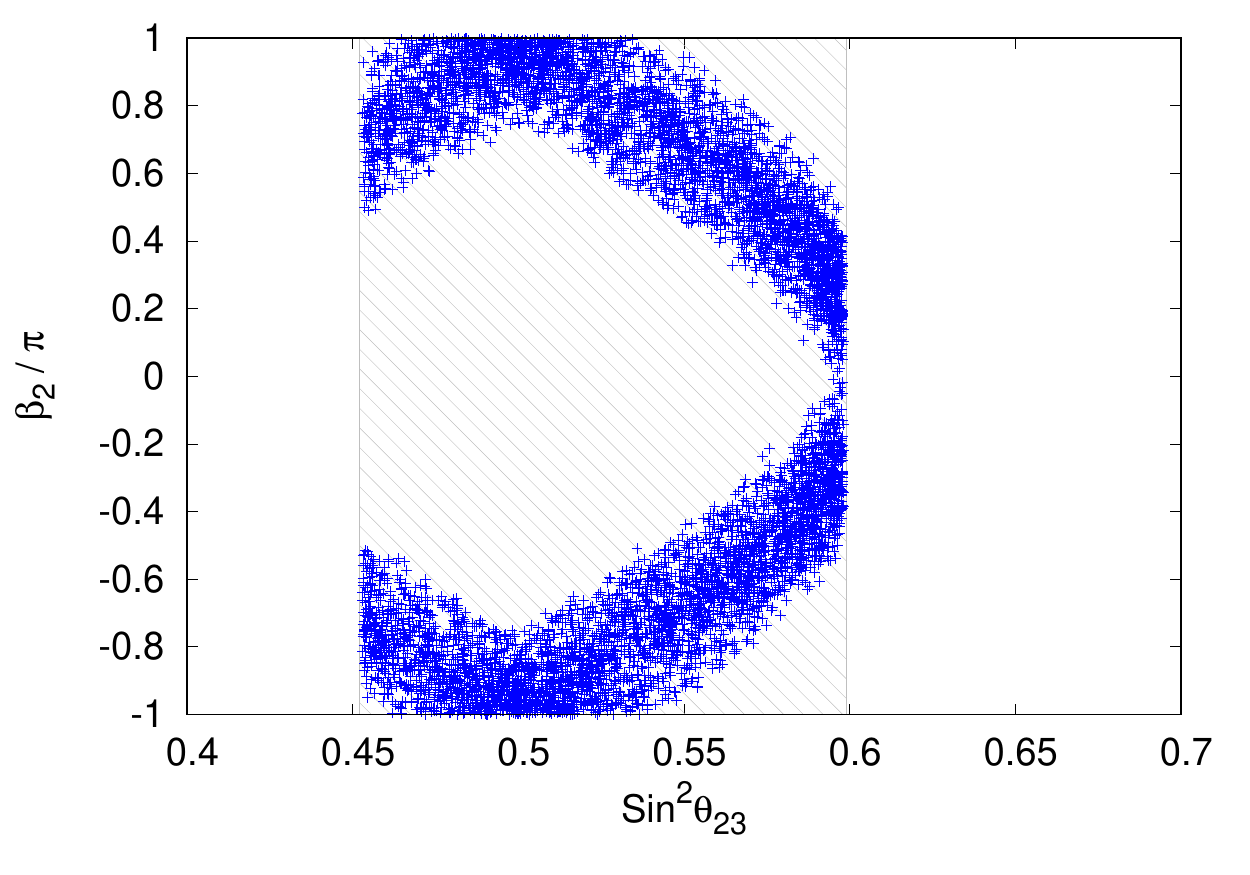}
\caption{Allowed region of $\beta_{2}$ Majorana phase in Case I for $\alpha_{1,2} \neq 0$. Blue (gray) region is obtained when the small $\mu-\tau$ symmetry breaking in the mass mtrix is considered (omitted). Correlation between $\beta_{2}$ and $\sin^2 \theta_{23}$ is shown for the $3\sigma$ range.}\label{fig:Majorananonzero}
\end{figure}

Now, we consider the case  $\alpha_{1}=0$. From Eq. (\ref{phases}), it is straightforward to show that this case directly implies that $\beta_1 = 0$,  and that $\beta_2$ is unbounded since $\alpha_2$ remains as a free parameter. Without considering small symmetry breaking requirements, this case does not give useful information about Majorana phases. Moreover, this case is totally ruled out when the requirement of a small symmetry breaking is imposed.

In the trivial case $\alpha_{1}=\alpha_2=0$, we also obtain $\beta_{1}=0$, while the correlation of $\delta_{CP}$ ($\beta_2$) with $\theta_{23}$ is given again by the gray region in Fig. \ref{fig:CP} (\ref{fig:Majorana}) when the symmetric limit is not included. Again, this case is completely disfavored when the requirement of a small symmetry breaking is added.

Finally, when we leave $\alpha_{1,2}$ as free parameters it is direct to show that $\beta_1$ remains free. The predicted region of $\delta_{CP}$ in this scenario is the same for both cases, when we include or omit the approximate symmetry condition, and coincides with the gray region in Fig. \ref{fig:CP}. The correlation between $\beta_2$ and $\sin^2 \theta_{23}$ is shown in Fig. \ref{fig:Majorananonzero}. Here, we can observe that the parameter space of $\beta_2$ is reduced when the small braking conditions are imposed (blue region). It is worth stressing that this scenario is the most favored by the experimental data when the small breaking requirements are included. 

Concerning to our Case I, we should stress that we have identified two possible scenarios predicting nice correlations between $CP$ violating parameters and the atmospheric mixing angle: the case  $\alpha_2=0$, and the general case where $\alpha_{1}$ and $\alpha_2$ remain free. Through these correlations, it would be possible to extract valuable information about $CP$ violation in the lepton sector in the advent of precise determinations of the mixing angles in forthcoming neutrino experiments. 

\begin{figure}
\includegraphics[scale=0.6]{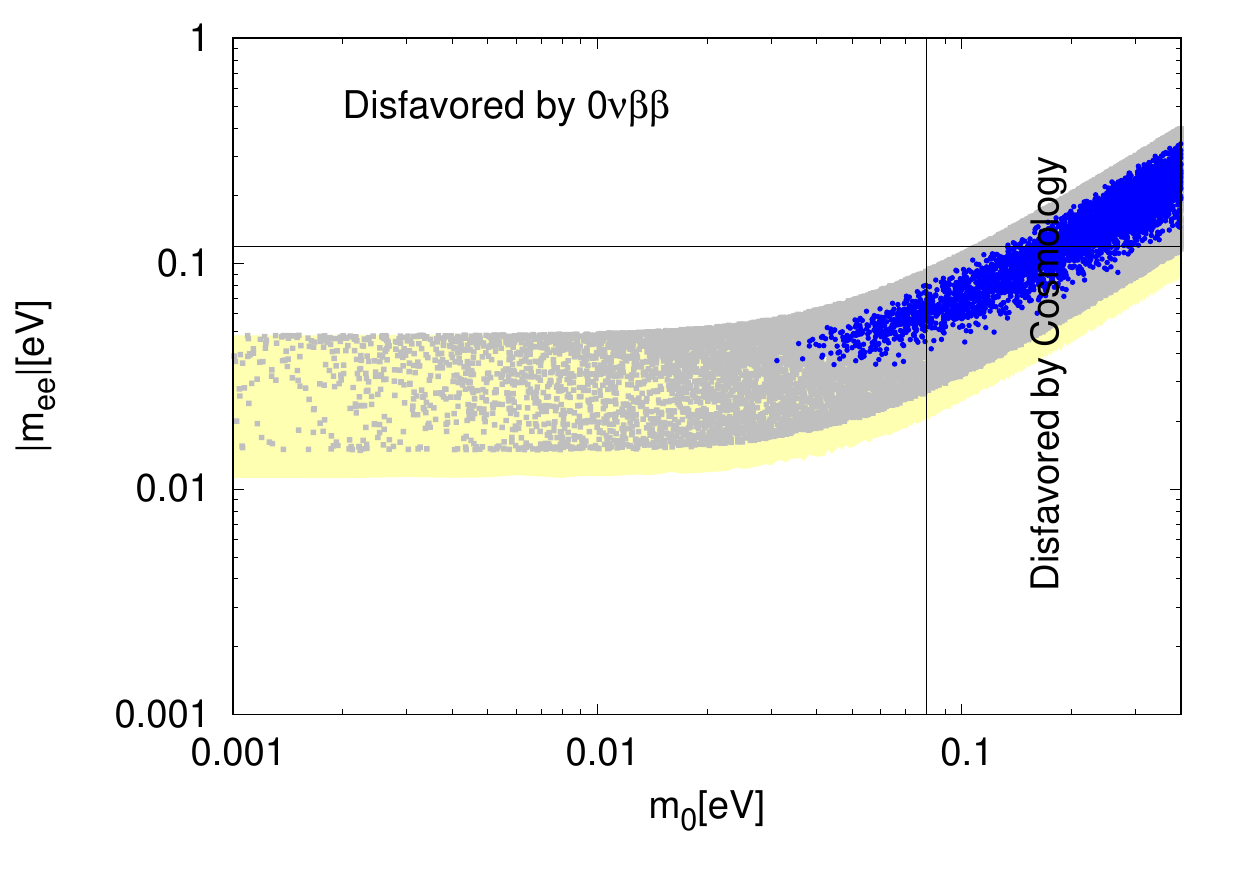}
\includegraphics[scale=0.6]{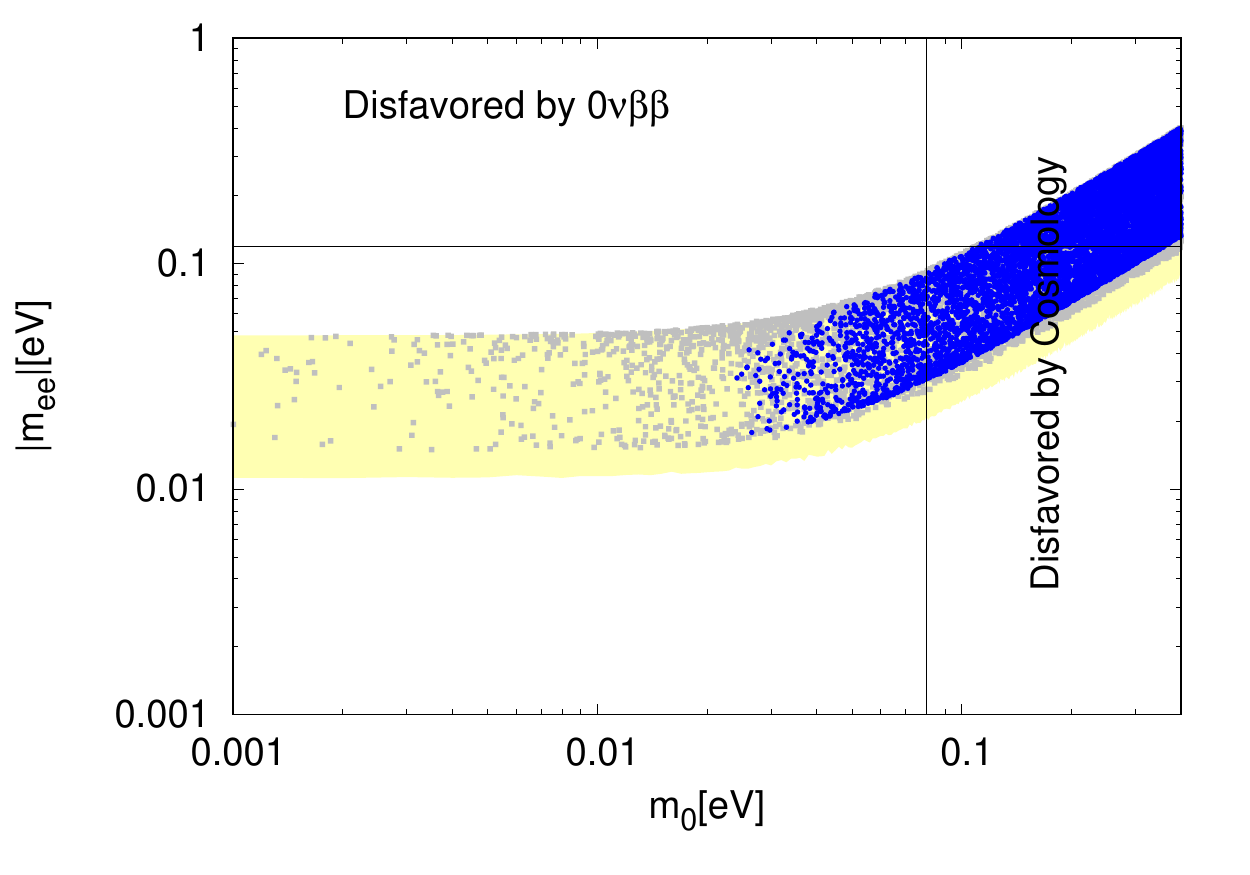}
\caption{$|m_{ee}|$ predicted regions in Case I. Left (right) plot corresponds to $\alpha_2=0$ ($\alpha_{1,2}\neq 0$). Scattered blue points (gray boxes) denote the allowed region when the approximate $\mu-\tau$ symmetry is (not) impossed. Yellow region is obtained for free Majorana phases.}\label{fig:0bb}
\end{figure}

Now, let us discuss some phenomenological implications of the different scenarios over the effective mass $|m_{ee}|$ describing neutrinoless double beta decay. In Fig. \ref{fig:0bb}, we show the predicted regions of $|m_{ee}|$ in the case of a 1-3 rotation for the inverted hierarchy. We plot separately the cases where $\alpha_{1,2}\neq 0$ (right) and $\alpha_2=0$ (left). In each case, we can observe how the region is reduced when the approximate symmetry requirement is included. As we have previously discussed, these are the only cases where at least one of the Majorana phases can be bounded. For comparison, we also plot the allowed region of the mass element $|m_{ee}|$ in the general case when both of the Majorana phases are free (yellow region). 

For $\alpha_2 = 0$ (left plot of Fig. \ref{fig:0bb}), and without including restrictions on the breaking parameters (gray boxes), the predicted region of $|m_{ee}|$ is slightly narrow compared to the case of totally free Majorana phases (yellow band). Owing to the correlation between $\beta_2$ and the atmospheric angle, this region could be narrowed even more with a more precise determination of $\theta_{23}$, as can be seen from Fig. \ref{fig:Majorana}. The implementation of small breakings in the mass matrix restricts the values of the lightest neutrino mass to lie above the 0.03 MeV, showing a marked preference for the quasi-degenerate hierarchy as was noted in \cite{Gupta:2013it,Rivera1,Rivera2}. This scenario is disfavored given its preference for maximal deviations in the atmospheric angle. 

On the other hand, in the general case $\alpha_1 , \alpha_2 \neq 0$, and when the small breaking condition is adopted (blue points), a strong preference for the quasi-degenerate hierarchy is also obtained (right plot of Fig. \ref{fig:0bb}). Hence, such scenario could be tested with improved limits of cosmology and/or neutrinoless double beta decay experiments. When the symmetry breaking condition is removed, the predicted region (gray boxes) is extended over the full range of $m_0$, as it is expected since both of the Majorana phases remains unbounded in this case.

\subsection{Case II: 2-3 rotation}

\begin{figure}
\includegraphics[scale=0.6]{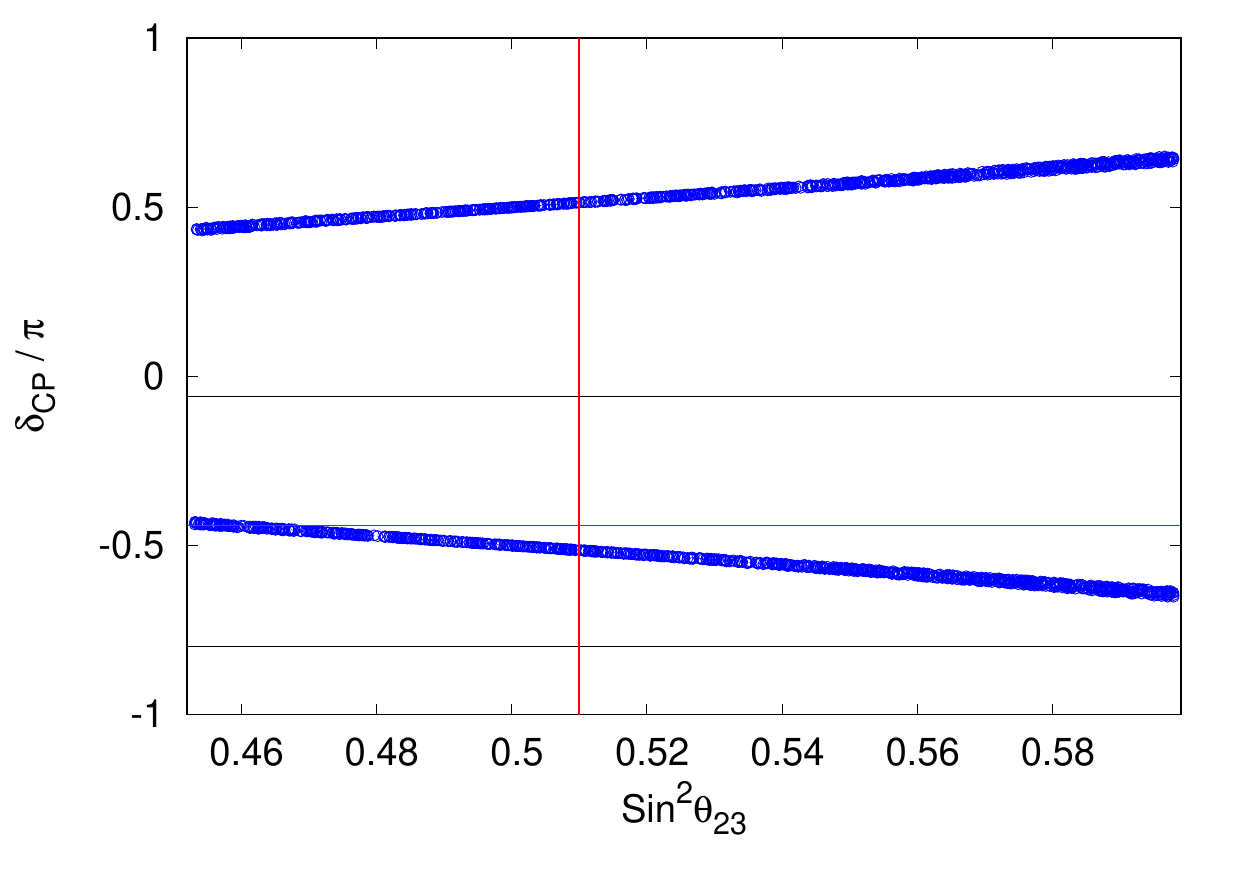} \\
\caption{Same description as in Fig. \ref{fig:CP} but for Case II.}\label{fig:CP2}
\end{figure}

Let us now move our attention to the case of a 2-3 rotation matrix. The selection criteria is similar to the one used in the previous case by allowing the predicted mixing angles to lie within the 3$\sigma$ experimental range. In Fig. \ref{fig:CP2} we show the predicted regions of $\delta_{CP}$ versus $\sin^2 \theta_{23}$ when we include (blue region) and omit (gray region) the requirements for small symmetry breaking. It should be noticed that blue and gray regions are overlapped, which means that the breaking parameters does not affect the parameters $\phi$ and $\sigma$ involved in predicting $\delta_{CP}$ (see Eq. \ref{CPphase2}).

\begin{figure}
\includegraphics[scale=0.6]{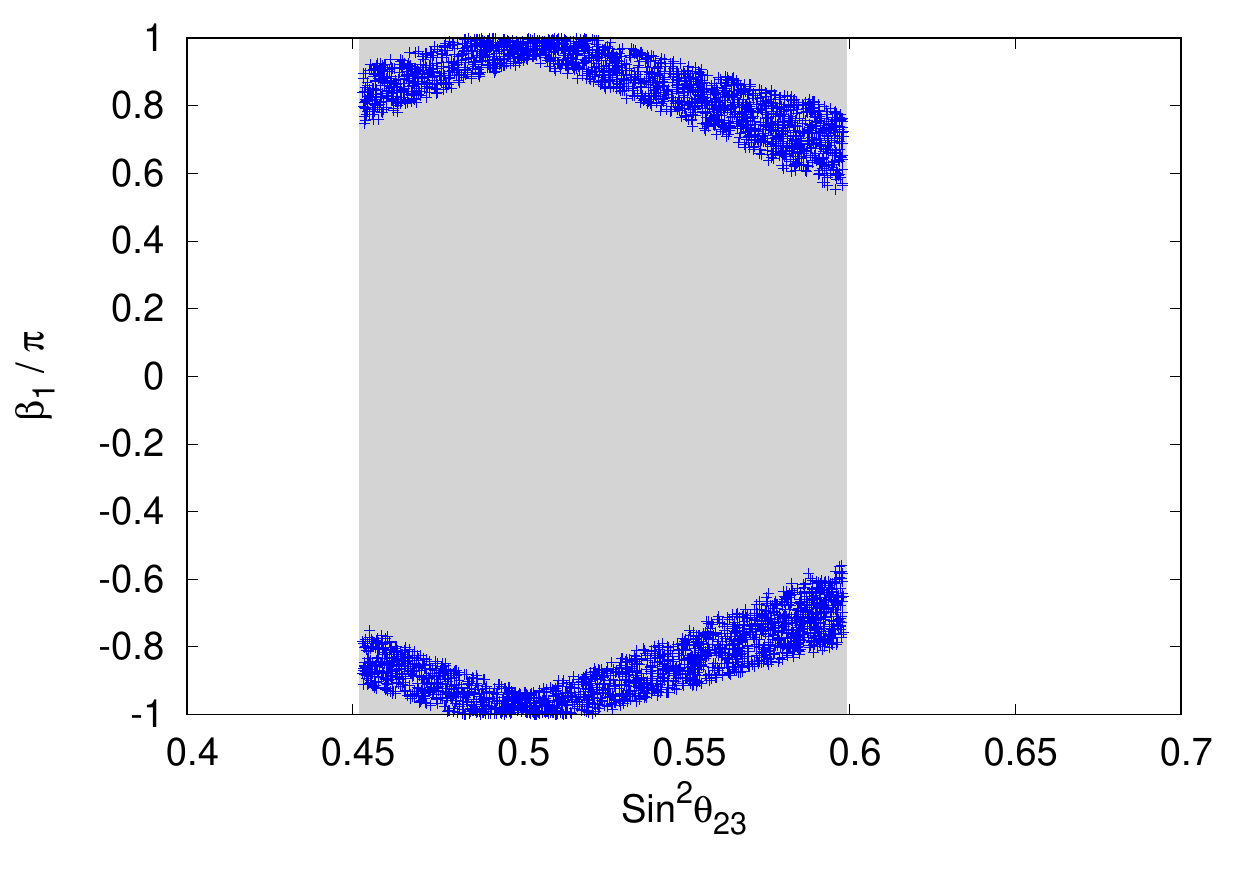}
\includegraphics[scale=0.6]{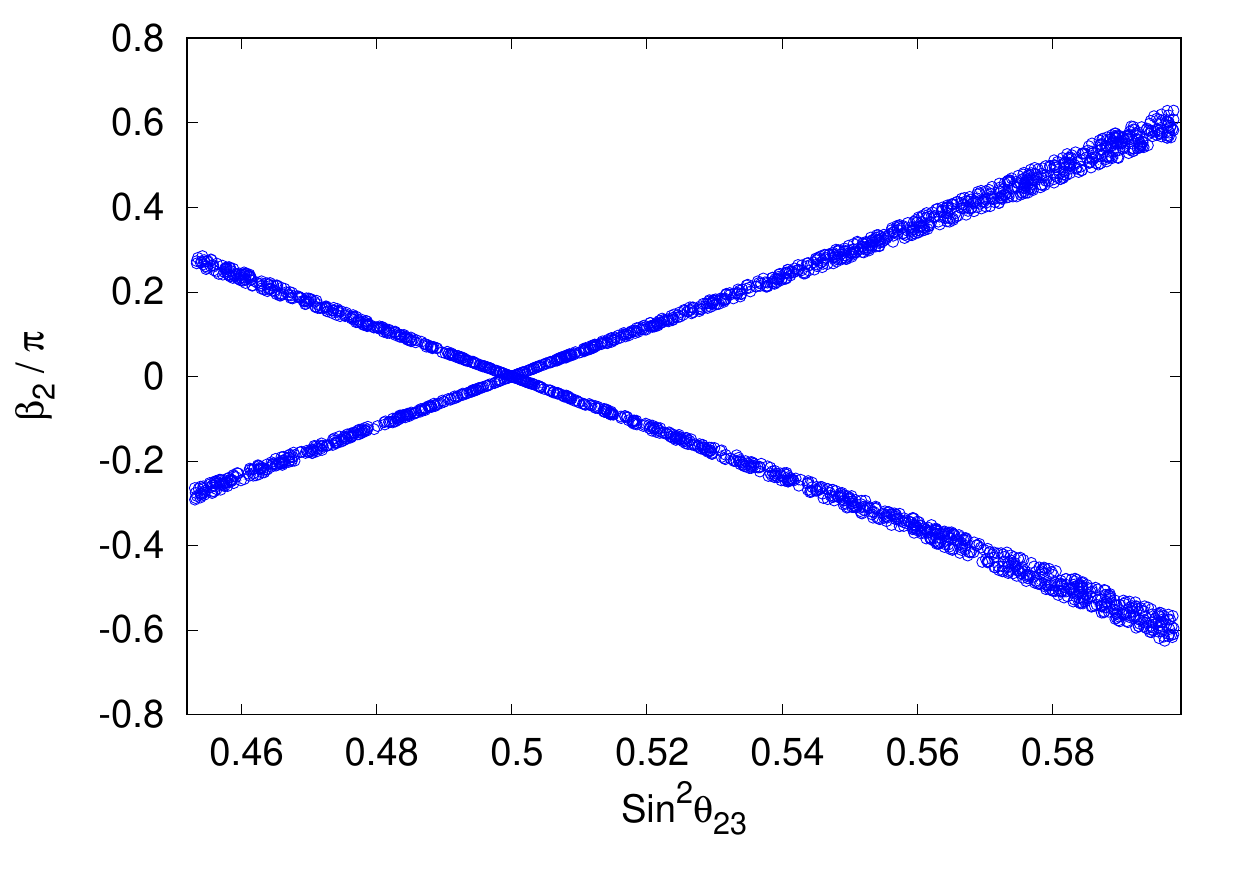}
\caption{Same description as in Fig. \ref{fig:Majorana} but for Case II.}\label{fig:Majorana2}
\end{figure}

 In Fig. \ref{fig:Majorana2}, we can see that the $\alpha_2=0$ scenario predicts well-defined regions of both Majorana $CP$ phases when the condition of an approximate symmetry is included (blue region). The omission of this last condition leaves $\beta_1$ undetermined (gray region in left plot) while $\beta_2$ remains unchanged (right plot), as it is expected due to the fact that $\beta_2$ does only depend on $\phi$ and $\sigma$ in this scenario.
 
 
 Other combinations of the phases $\alpha_1$ and $\alpha_2$ leave Majorana phases undetermined, or fixed to zero, which give no relevant information even if we ask for small symmetry breakings in the neutrino mass matrix. 
 
\begin{figure}
\includegraphics[scale=0.6]{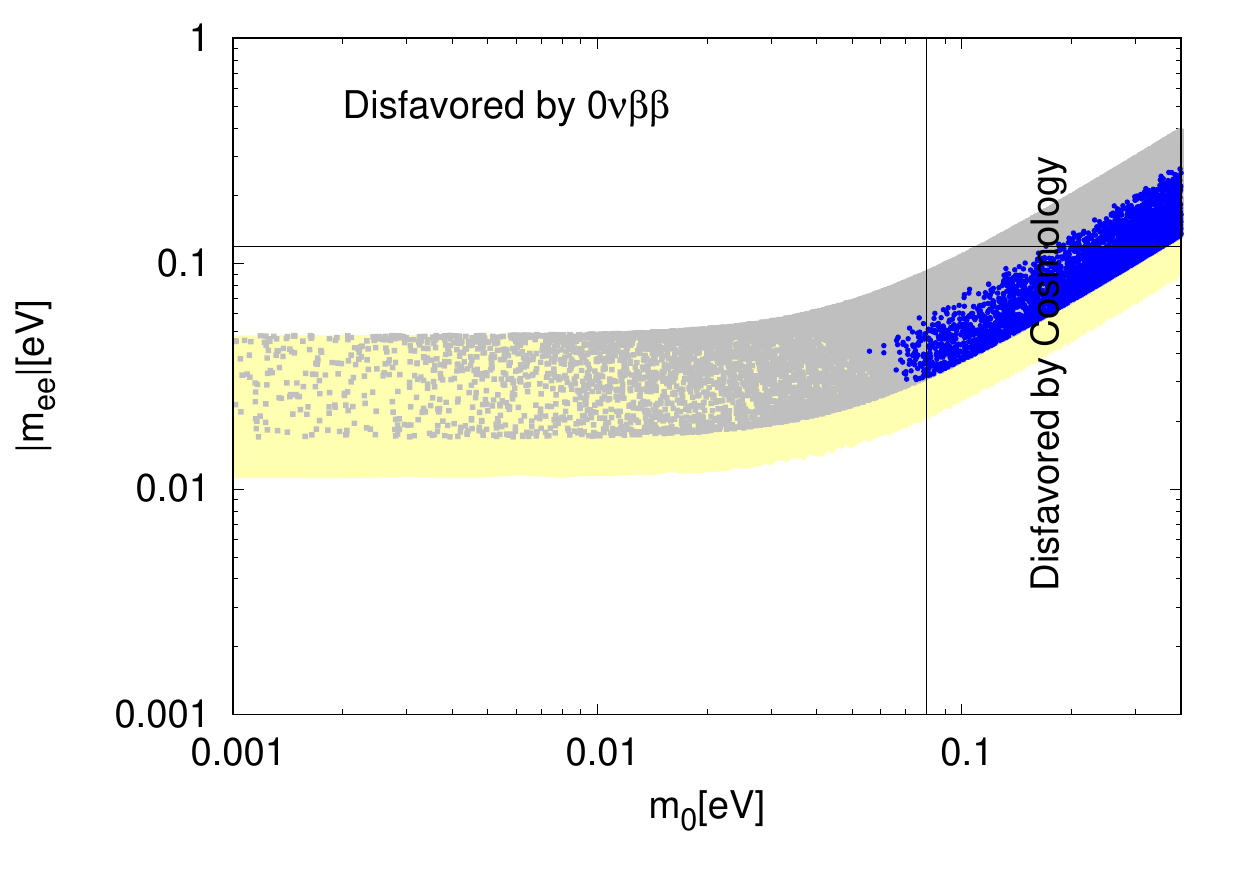}
\caption{Same description as in Fig. \ref{fig:0bb} but for Case II.}\label{fig:0bb2}
\end{figure}

 The predicted region of $|m_{ee}|$ is shown in Fig. \ref{fig:0bb2}, which corresponds to the case $\alpha_2=0$. The scattered gray boxes (blue points) show the predicted values of $|m_{ee}|$ when the small symmetry breaking condition is omitted (included). As we have discussed, the case $\alpha_2=0$ with the small breaking requirement is the only one where both Majorana phases can be bounded. This is reflected in a narrow region of $|m_{ee}|$ (blue points in Fig. \ref{fig:0bb2}), which is disfavored by the current cosmological limit of $m_{0}$. 
 

\section{Summary}

It is well known that the TBM mixing pattern predicts maximal value for the atmospheric angle ($\theta_{23}=\pi/4$) and null $\theta_{13}$. The latter value is disfavored by recent experimental data while the former is still allowed. Despite the confirmation of a sizable reactor angle, the TBM pattern remains as a simple alternative to explain neutrino mixings if some modifications are adopted. 

 In this study, we have parametrized the neutrino mixing matrix by considering deviations from the TBM pattern through a correction matrix in the neutrino sector, which we wrote in terms of four correction parameters. Given its comparison with the standard parametrization, we obtained analytic expressions for the mixing angles, the Dirac $\delta_{CP}$, and the Majorana phases, in terms of these parameters. We have analyzed, in addition, two breaking parameters which helped to define an approximate $\mu-\tau$ symmetry in the mass matrix. These breaking parameters could also be written in terms of the correction parameters and played an important role in bounding the correction phases. In total, we have considered five restrictions (three from the experimental values of the mixing angles, and two from the symmetric limit in the mass matrix) in order to bound the correction parameters, and hence to predict the allowed regions for Majorana and Dirac $CP$ phases in this context. 
 
 
 Concerning our scheme, the predicted $CP$ phases could be related with the atmospheric angle in different scenarios. We have also analyzed the phenomenological implications of the predicted Majorana phases over the effective mass $|m_{ee}|$ describing neutrinoless double beta decay. In this line of thought, our results showed marked preference for the quasi-degenerate hierarchy when the approximate $\mu-\tau$ symmetry requirement was included. Future improvements in the determination of the mixing angles and the Dirac $CP$ phase, joint to the effective mass $|m_{ee}|$, will play a fundamental role in testing such scenarios.


\section*{Acknowledgements}

SLT acknowledge Universidad Santiago de Cali for the hospitality. 

\bibliographystyle{utphys}
 
\bibliography{Mu-Tau.bbl}
   

\end{document}